\documentclass[aps,amsmath,amssymb,twocolumn,floatfix,superscriptaddress]{revtex4-2}
\pdfoutput=1
\usepackage[dvipdfmx]{graphicx}
\usepackage[justification=Justified,
labelfont=bf,
format=plain]{caption}
\usepackage{ragged2e}
\usepackage{dcolumn}
\usepackage{bm}
\usepackage{float}
\usepackage{subfig}
\usepackage{enumitem}
\usepackage[colorlinks=true,urlcolor=blue,citecolor=blue,linkcolor=blue]{hyperref}
\usepackage{amsmath}
\usepackage{amsfonts}
\usepackage{amssymb}
\usepackage{amsthm}
\usepackage{mathtools}
\usepackage{bm}
\usepackage[capitalise]{cleveref}
\usepackage{physics}
\usepackage{xparse}
\usepackage{url}
\newcommand{\sign}{\text{sgn}}
\usepackage{textcomp}
\newcommand{\texttildecenter}{\raisebox{0.5ex}{\texttildelow}}

\makeatletter
\newsavebox{\@brx}
\newcommand{\llangle}[1][]{\savebox{\@brx}{\(\m@th{#1\langle}\)}
  \mathopen{\copy\@brx\kern-0.5\wd\@brx\usebox{\@brx}}}
\newcommand{\rrangle}[1][]{\savebox{\@brx}{\(\m@th{#1\rangle}\)}
  \mathclose{\copy\@brx\kern-0.5\wd\@brx\usebox{\@brx}}}
\makeatother

\newcommand{\kket}[1]{
	\ensuremath{|{#1}\rrangle}
}
\newcommand{\bbra}[1]{
	\ensuremath{\llangle{#1}|}
}
\newcommand{\bbraket}[1]{
	\ensuremath{\llangle{#1}\rrangle}
}

\usepackage{xpatch}
\xpatchbibmacro{textcite}{\addspace}{\addnbspace}{}{}

\crefname{figure}{Fig.}{Figs.}
\crefname{equation}{Eq.}{Eqs.}
\crefname{section}{Sec.}{Sec.}
\Crefname{figure}{Figure}{Figures}
\Crefname{equation}{Equation}{Equations}
\Crefname{section}{Section}{Sections}

\crefformat{equation}{Eq.~\!#2(#1)#3}
\crefformat{figure}{Fig.~\!#2#1#3}
\Crefformat{equation}{Equation~\!#2(#1)#3}
\Crefformat{figure}{Figure~\!#2#1#3}
\crefmultiformat{figure}{Figs.~(#2#1#3)}
{ and~(#2#1#3)}{, (#2#1#3)}{ and~(#2#1#3)}
\crefmultiformat{figure}{Figs~#2#1#3}{ and~#2#1#3}{, #2#1#3}{, and~#2#1#3}

\newcommand{\crefSubFigRef}[2]{\crefformat{figure}{Fig.~\!##2##1{(#2)}##3}%
  \cref{#1}\crefformat{figure}{Fig.~\!##2##1##3}}

\newcommand{\crefMultiSubFigRef}[2]{\crefformat{figure}{Figs.~\!##2##1{(#2)}##3}%
  \cref{#1}\crefformat{figure}{Fig.~\!##2##1##3}}
  
\newcommand{\CrefSubFigRef}[2]{\Crefformat{figure}{Figure~##2##1{(#2)}##3}%
  \Cref{#1}\Crefformat{figure}{Figure~##2##1##3}}

\begin{document}
\preprint{APS/123-QED}
\ExplSyntaxOn
\NewDocumentCommand{\convertto}{mm}
 {
  \texttt{#2~=~\fp_to_decimal:n { round ( (#2)/(1#1), 5 ) }#1}
 }
\DeclareExpandableDocumentCommand{\thelength}{ O{mm} m }
 {
  \fp_to_decimal:n { round ( #2/1#1, 5 ) } #1
 }
\ExplSyntaxOff

\title{Experimental demonstration of a high-fidelity virtual two-qubit gate}

\author{Akhil Pratap Singh}
\affiliation{Department of Applied Physics, Graduate School of Engineering,
The University of Tokyo, Bunkyo-ku, Tokyo, 113-8656, Japan}
 
\author{Kosuke Mitarai}
\affiliation{Graduate School of Engineering Science, Osaka University
1-3 Machikaneyama, Toyonaka, Osaka 560-8531, Japan} 

\author{Yasunari Suzuki}
\affiliation{NTT Computer and Data Science Laboratories, NTT Corporation, Musashino 180-8585, Japan}

\author{Kentaro~Heya}
\affiliation{RIKEN Center for Quantum Computing (RQC), Wako, Saitama 351-0198, Japan}

\author{Yutaka Tabuchi}
\affiliation{RIKEN Center for Quantum Computing (RQC), Wako, Saitama 351-0198, Japan}

\author{Keisuke Fujii}
\affiliation{Graduate School of Engineering Science, Osaka University
1-3 Machikaneyama, Toyonaka, Osaka 560-8531, Japan}
\affiliation{RIKEN Center for Quantum Computing (RQC), Wako, Saitama 351-0198, Japan}

\author{Yasunobu Nakamura}
\affiliation{Department of Applied Physics, Graduate School of Engineering,
The University of Tokyo, Bunkyo-ku, Tokyo, 113-8656, Japan}
\affiliation{RIKEN Center for Quantum Computing (RQC), Wako, Saitama 351-0198, Japan}

\begin{abstract}    
We experimentally demonstrate a virtual two-qubit gate and characterize it using quantum process tomography~(QPT). The virtual two-qubit gate decomposes an actual two-qubit gate into single-qubit unitary gates and projection gates in quantum circuits for expectation-value estimation. We implement projection gates via mid-circuit measurements. The deterministic sampling scheme reduces the number of experimental circuit evaluations required for decomposing a virtual two-qubit gate. We also apply quantum error mitigation to suppress the effect of measurement errors and improve the average gate fidelity of a virtual controlled-$Z$ (CZ) gate to $f_{\rm av} = 0.9938 \pm 0.0002$. Our results highlight a practical approach to implement virtual two-qubit gates with high fidelities, which are useful for simulating quantum circuits using fewer qubits and implementing two-qubit gates on a distant pair of qubits.
\end{abstract}

\pacs{Valid PACS number}
\maketitle

\section{\label{sec:Introduction}INTRODUCTION}
Quantum computing research is progressing on a promising yet challenging path to realize a scalable fault-tolerant quantum computer, which is expected to have the capabilities to solve many problems intractable for their classical counterparts. However, current quantum devices with limited coherence times, low scalability, and non-negligible noises, termed as noisy intermediate scale quantum, or NISQ devices~\cite{Preskill2018quantumcomputingin}, are still far from being a full-fledged quantum computer. Nevertheless, they are still proving to be the testbeds for many promising quantum algorithms and quantum applications~\cite{nielsen_chuang_2010,nisq_review_bharti2022,Peruzzo_vqe_2014,MonteCarlo_Ashley,vqe_review_Cerezo2021,quantum_chemistry_Cao2019,quantum_chemistry_Endo_2020}. The recent experimental demonstrations of quantum computation with over 50 qubits~\cite{google_supremacy_Arute2019,Yulin_Quantum_Advantage_PRL,Zhu_Quantum_Advantage_ScienceBulletin,Kim_IBM127_errormitigation_Nature2023} have motivated the practical interest in solving large problems using smaller quantum devices, despite the trade-off of utilizing more classical resources.

To maximize the capabilities of a limited-sized NISQ quantum processor, various techniques for simulating large quantum circuits with smaller quantum devices have been proposed and demonstrated~\cite{Bravyi_PRX_2016,Peng_PRL_2020,Chong_exp_PRL,Eddins_PRXQuantum_EntanglementForging,CircuitKnitting_IBM,circuit_knitting_optimal,decomposition_Fujii_PRXQuantum2022,Endo_PRX_2018}. These techniques are useful for the expectation value estimation of a large quantum circuit, as they can reduce the hardware requirement by ``cutting" the circuit, albeit with some overhead cost. 
\textcite{mitarai_virtual_gate} proposed the ``virtual two-qubit gate" technique, a general decomposition scheme for a two-qubit gate. The virtual two-qubit gate allows us to simulate a two-qubit gate from a quasi-probability decomposition of local single-qubit operations in the quantum circuits used for the expectation value estimation for observables. The virtual two-qubit gate scheme has been experimentally utilized on a distant pair of superconducting qubits to reduce the number of SWAP operations required and thus reducing the two-qubit errors~\cite{virtual_gate_ibm}. However, the characterization of the virtual two-qubit gate was not performed, thereby limiting the ability to evaluate its quality.

In this work, we experimentally demonstrate a virtual two-qubit gate taking the example of a controlled-$Z$~(CZ) gate and characterize it through the quantum process tomography~(QPT)~\cite{Chow_RB_QPT_PRL}. The virtual two-qubit gate requires the implementation of projection gates, which are non-unitary. We implement the projection gates through mid-circuit measurements. However, this limits the fidelity of the virtual two-qubit gate since measurement errors are typically higher than single-qubit gate errors. We thus formulate the quantum error mitigation for mid-circuit measurements and apply it to improve the average gate fidelity of the virtual two-qubit gate.

The rest of the paper is organized as follows: In~\cref{sec:methods}, we review the virtual two-qubit gate decomposition technique, and present our approach for its experimental implementation. We then discuss combining the quantum error mitigation with the virtual two-qubit gate. \Cref{sec:experiment} provides details on the experimental device, gate implementation, and the characterization of the virtual CZ gate. We also demonstrate the improved average gate fidelity of the virtual CZ gate after incorporating the quantum error mitigation. Lastly, in~\cref{sec:Conclusion_Discussion} we summarize our work and discuss potential directions for future work.

\section{\label{sec:methods}Methods}
\begin{figure*}[t]
\includegraphics[width=\textwidth,keepaspectratio]{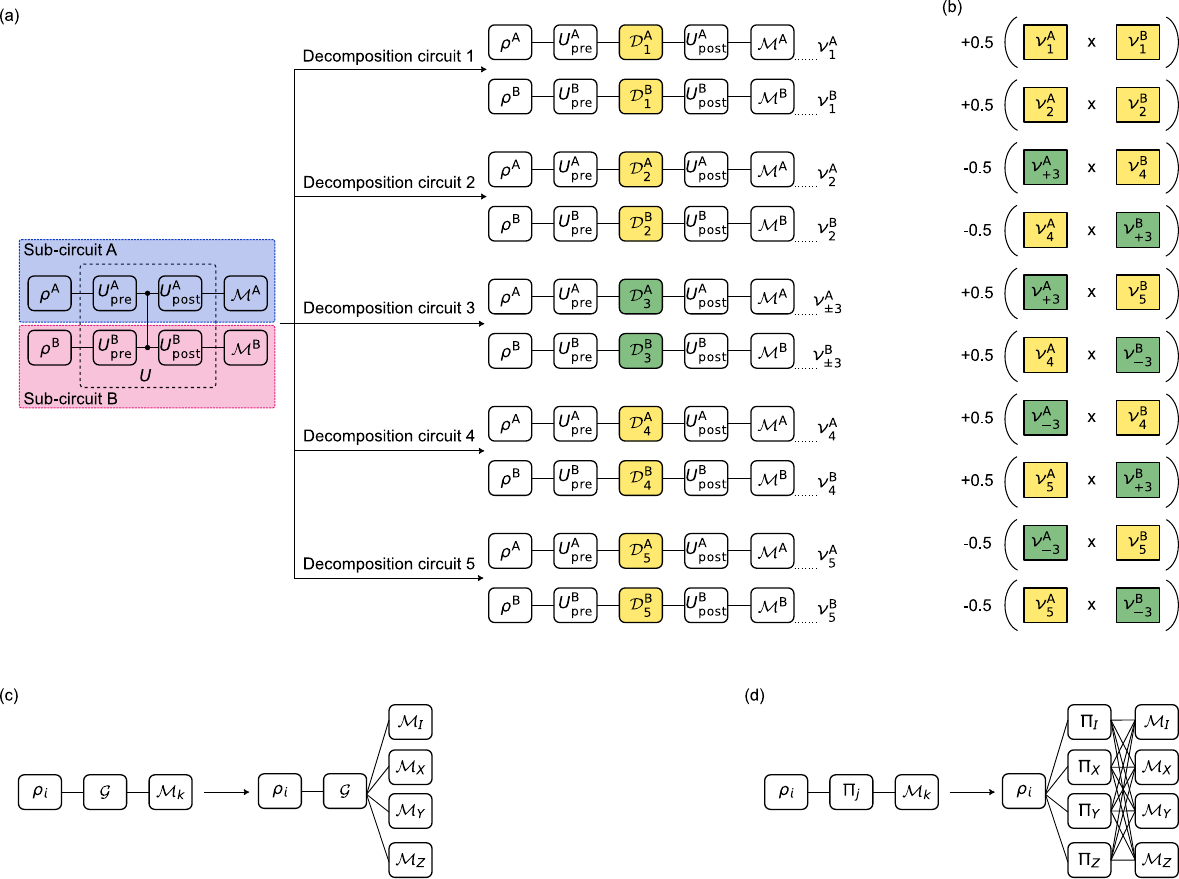}
\caption{\label{fig:CZ_decomposition}\textbf{Decomposition of a CZ gate and quantum error mitigation.} (a)~Decomposition of a simple two-qubit quantum circuit with a single CZ gate and other one-qubit gates. See text for the explanation. (b)~Calculation of the final expectation value using the expectation values $\nu_{j}^{i}$ obtained in the respective decomposition circuits in~(a). (c)~Measurement error mitigation for a single-qubit unitary operation. (d)~Quantum error mitigation for a projection gate and a measurement.}
\end{figure*}

\subsection{\label{sub_sec:gate_decomposition}Gate Decomposition}
We can break down any two-qubit gate expressed in the form $e^{i \theta A \otimes B}$ into six local single-qubit operations, where $A^{2} = B^{2}=I$,  and $I$ is an identity matrix. We use the tilde~(\texttildecenter) symbol to represent a superoperator~$\Tilde{U}$ corresponding to an operator~$U$ whose action on a density matrix $\rho$ is defined as $\Tilde{U} \rho = U \rho U^{\dagger}$.

The decomposition of a CZ gate is written as
\begin{align}
\widetilde{\mathrm{CZ}} & =  \frac{1}{2}
\biggl(
    \Tilde{\mathcal{R}}_{Z} \left(\frac{\pi}{2} \right) \otimes \Tilde{\mathcal{R}}_{Z} \left(\frac{\pi}{2} \right)\biggr) \nonumber \\
& {} +
\frac{1}{2} 
\biggl(
   \Tilde{\mathcal{R}}_{Z} \left(-\frac{\pi}{2} \right) \otimes \Tilde{\mathcal{R}}_{Z} \left(-\frac{\pi}{2} \right)\biggr) \nonumber \\
& {} -
\frac{1}{2}  \sum_{\alpha_{1},\alpha_{2}} \alpha_{1} \alpha_{2}
\bigg[
        \Tilde{\Pi}_{\alpha_{1}Z} \otimes \Tilde{\mathcal{R}}_{Z} \left((\alpha_{2}+1)\frac{\pi}{2}\right)
    \nonumber \\
& \qquad \qquad + 
        \Tilde{\mathcal{R}}_{Z} \left((\alpha_{1}+1)\frac{\pi}{2}\right) \otimes \Tilde{\Pi}_{\alpha_{2}Z} 
        \bigg] \;,
\label{eq:CZ_full_decomp}
\end{align}
where $\alpha_{1}, \alpha_{2} \in \{-1, +1 \}$. $\Tilde{\mathcal{R}}_{Z}\qty(\pm \theta)$ and $\Tilde{\Pi}_{\pm Z}$ are superoperators corresponding to the operators $\mathcal{R}_{Z}\qty(\pm \theta) = e^{\pm i Z \theta/2}$ and $ \Pi_{\pm Z}=(I\pm Z)/2 $, respectively. $\Pi_{\pm Z}$ are the (non-destructive) projection gates, which project the quantum state onto the respective eigenbasis and are non-unitary~\cite{mitarai_virtual_gate}.

Reference~\!\cite{mitarai_virtual_gate} has quantified the decomposition cost in terms of quasi-probability simulation where the decomposition operators in \cref{eq:CZ_full_decomp} are sampled with probabilities proportional to their coefficients~\cite{Hakkaku2022}. The set of these decomposition operators can also be written as
\begin{align}
    \mathcal{D}_{\rm CZ} =& \bigg\{\mathcal{R}_Z\qty(\frac{\pi}{2})\otimes \mathcal{R}_Z\qty(\frac{\pi}{2}), 
    \mathcal{R}_Z\qty(-\frac{\pi}{2})\otimes \mathcal{R}_Z\qty(-\frac{\pi}{2}), \nonumber \\ 
    & \mathcal{R}_Z\qty(\pi)\otimes \Pi_{\pm Z}, 
    I\otimes \Pi_{\pm Z}, \Pi_{\pm Z}\otimes \mathcal{R}_Z(\pi), \nonumber \\
    & \Pi_{\pm Z} \otimes I\bigg\} \;.
    \label{eq:CZ_decomp_set}
\end{align}

In this work, we employ deterministic sampling for the decomposition operators, which means that we perform uniform experiments with each operator from~\cref{eq:CZ_decomp_set} in a deterministic manner. Unlike the quasi-probability approach, we uniformly sample the decomposition operators with a probability of unity, regardless of their coefficients. We do that because of following two reasons. First, by implementing deterministic sampling, we can reduce the total number of circuit evaluations for each sub-circuit from six to five. It is explained in detail later in this section. Second, apart from the virtual two-qubit gate implementation, we also perform its characterization. To obtain the process matrix for the virtual CZ gate, we need process matrices of all the decomposition operators $\mathcal{D}^{i}_{j}$~(see~\cref{subsec:qpt}). Thus, we implement each decomposition operation deterministically.

\CrefSubFigRef{fig:CZ_decomposition}{a} illustrates the circuit decomposition for a virtual CZ gate in terms of experimental implementation. We consider a simple two-qubit quantum circuit consisting of single-qubit gates and a CZ gate. The aim here is to do the expectation value estimation for this two-qubit circuit using only one qubit by replacing an actual CZ gate with a virtual CZ gate. For the virtual two-qubit gate decomposition, we cut the CZ gate and consider the remaining circuit consisting of two one-qubit sub-circuits A and B. Each sub-circuit consists of the initial state $\rho^{i}$, set of single-qubit gates implemented before and after the CZ gates denoted by $U^{i}_{\rm pre}$ and $U^{i}_{\rm post}$, respectively, and the measurement of observable $k$, $\mathcal{M}^{i}_{k}$, where the superscript $i \in \{\rm A,B\}$ refers to the sub-circuit they belong to.
In order to do the expectation value estimation using a virtual two-qubit gate, we do five different circuit evaluations, labelled as ``Decomposition circuits" in~\crefSubFigRef{fig:CZ_decomposition}{a}, where we replace the actual CZ gate with local decomposition operators $\mathcal{D}^{i}_{j}$ on both control and target qubits shown on the right hand side of~\crefSubFigRef{fig:CZ_decomposition}{a}.
In $\mathcal{D}^{i}_{j}$, $i \in \{\rm A,B\}$ and $j \in \{1,2,3,4,5 \}$ refer to the indices of the sub-circuits and the decomposition circuits, respectively.
Single-qubit unitary operations $\mathcal{D}_{j=1,2,4,5}^{i} \in \{\mathcal{R}_{Z}\qty(\pi/2),\mathcal{R}_{Z}\qty(-\pi/2), I, Z\}$ are shown in the yellow boxes, while projection gates $\mathcal{D}_{j=3}^{i} = \{\Pi_{\pm Z}\}$ are shown in the green boxes. For the expectation value estimation using the virtual two-qubit gate, we substitute the expectation values $\nu^{i}_j$ obtained from decomposition circuits~1--5 in~\cref{eq:CZ_full_decomp} and get the final expectation value. This calculation is illustrated in~\crefSubFigRef{fig:CZ_decomposition}{b}, which is a graphical representation of~\cref{eq:CZ_full_decomp}, where the ten pairs of boxes correspond to the ten terms in~\cref{eq:CZ_full_decomp}.

Note that the utilization of the virtual two-qubit gate can allow us to decompose a quantum circuit $U$ into two completely separate quantum circuits each acting only on qubits in sub-circuits A and B, as considered in~\crefSubFigRef{fig:CZ_decomposition}{a}. We take input to the circuit as a separable state $\rho = \rho_{\rm A}\otimes \rho_{\rm B}$. As an output, we wish to obtain expectation value $P=P_{\rm A}\otimes P_{\rm B}$, where $P_{\rm A}$ and $P_{\rm B}$ are Pauli operators acting on groups A and B, respectively. For example, if we define two superoperators $\Tilde{U_1}$ and $\Tilde{U_2}$ then their tensor product is expressed as  
\begin{align}
    ( \Tilde{U_1}\otimes\Tilde{U_1}) \rho = \qty(U_1\otimes U_2) \rho (U_1^\dagger\otimes U_2^\dagger) \;.
    \label{eq:quantum_map_tensored}
\end{align}
Using \cref{eq:quantum_map_tensored}, $U$, as shown in~\crefSubFigRef{fig:CZ_decomposition}{a}, can be decomposed as

\begin{align}
    \Tilde{U} = \sum_j &\qty(\Tilde{U}_{\rm post}^{\rm A}\otimes \Tilde{U}_{\rm post}^{\rm B}) \qty(\Tilde{\mathcal{D}}_j^{\rm A}\otimes\Tilde{\mathcal{D}}_j^{\rm B}) \nonumber \\
    & \qty(\Tilde{U}_{\rm pre}^{\rm A}\otimes \Tilde{U}_{\rm pre}^{\rm B}) \;.
\end{align}

The desired expectation value can be written as
\begin{align}
    \Tr\qty(P\Tilde{U}\rho)  = 
    &\sum_j \bigg[\Tr\qty(P_{\rm A}\Tilde{U}_{\rm post}^{\rm A}\Tilde{\mathcal{D}}_j^{\rm A} \Tilde{U}_{\rm pre}^{\rm A} \rho_{\rm A}) \nonumber \\
    &\Tr\qty(P_{\rm B}\Tilde{U}_{\rm post}^{\rm B}\Tilde{\mathcal{D}}_j^{\rm B} \Tilde{U}_{\rm pre}^{\rm B} \rho_{\rm B}) \bigg] \;.
    \label{eq:tensored_expVal}
\end{align}

Therefore, in this case, we can estimate all of the values of $\Tr\qty(P_{\rm A}\Tilde{U}_{\rm post}^{\rm A}\Tilde{\mathcal{D}}_j^{\rm A} \Tilde{U}_{\rm pre}^{\rm A} \rho_{\rm A})$ and $\Tr\qty(P_{\rm B}\Tilde{U}_{\rm post}^{\rm B}\Tilde{\mathcal{D}}_j^{\rm B} \Tilde{U}_{\rm pre}^{\rm B} \rho_{\rm B})$ separately for $j=1,\ldots, 5$~(as shown in~\crefSubFigRef{fig:CZ_decomposition}{a}) and then combine them according to~\cref{eq:CZ_full_decomp} and ~\cref{eq:tensored_expVal} to obtain the result for $U$. We note that for quasi-probability sampling of the decomposition operations, \cref{eq:CZ_full_decomp} needs six different circuit evaluations because there are six distinct operations (see \cref{eq:CZ_decomp_set}) involved in~\cref{eq:CZ_full_decomp}. However, by comparing~\cref{eq:CZ_full_decomp}, \cref{eq:CZ_decomp_set}, and \cref{eq:tensored_expVal}, we see that if we sample the decomposition operators $\mathcal{D}_{j}^{i}$ deterministically then we can reduce the total number of circuit evaluations for each sub-circuit from six to five. This can be done since, by implementing the projection gates through mid-circuit measurements, we obtain both expectation values of the projection operators ($\Pi_{\pm Z}$ in case of a CZ gate) in a single experiment ($\nu_{\pm 3}^{\rm A,B}$ in~\crefSubFigRef{fig:CZ_decomposition}{a}).

\subsection{\label{measurement_error_mitigation} Combining quantum error mitigation with the virtual two-qubit gate}
We use mid-circuit measurements to implement the projection gates. The measurement errors become the dominant contributor to the average gate infidelity of the virtual two-qubit gate since the measurement errors~($3.9\%$) are much larger than the initialization errors~($\sim$1\%) and single-qubit errors~($<0.05\%$) in our device. The details of the device are presented in~\cref{sub_sec:system_details}. In order to mitigate the measurement errors, we perform quantum error mitigation~(QEM) with the probabilistic error cancellation~(PEC) method~\cite{Endo_PRX_2018,QPMEM_science}. In this method, we define a set of modified circuits for the original circuit in which we wish to mitigate the errors. Then, the original circuit is replaced randomly by a selected set of modified circuits, where the probabilities to select the random circuits are calculated before running the quantum circuits. The mitigated result is obtained by taking the average of the modified circuits. Here, we apply this error mitigation technique both to the mid-circuit measurements for the projection gates and to the final measurements. \CrefSubFigRef{fig:CZ_decomposition}{c} illustrates the measurement error mitigation circuit for single-qubit unitary operations where $\mathcal{G} = \mathcal{D}_{j=\{1,2,4,5\}}$, and instead of the measurement of the desired observable $k$, $\mathcal{M}_k$, we measure the set of Pauli operators denoted as $\mathcal{M}_{\mathcal{P}} = \mathcal{M}_{\qty{I,X,Y,Z}}$. In the experiments, measurements on the $X$ and $Y$ basis are implemented by applying $\mathcal{R}_{Y}(\pi/2)$ and $\mathcal{R}_{X}(\pi/2)$ gates, respectively, just before the usual $Z$ basis measurement. For measurement in the $I$ basis, we always consider the expectation value of operator $I$ to be unity. In case of the projection gates $\mathcal{D}_{j=3}$, we apply the probabilistic quantum error mitigation sequentially~\cite{QPMEM_science}, first for the projection gate $\Pi_{j}$, implemented through mid-circuit measurement, and then on the final measurement $\mathcal{M}_k$ as shown in \crefSubFigRef{fig:CZ_decomposition}{d}. See \Cref{app_sec:mem} for the details.

\section{\label{sec:experiment}Experiment}
\subsection{\label{sub_sec:system_details}System details}
Local single-qubit operations can be classified into two categories: unitary and non-unitary/projection gates. To implement a virtual CZ gate experimentally, we use a fixed-frequency superconducting transmon qubit~\cite{Koch_transmon}. The qubit is a part of a 16-qubit device~\cite{tamate2021scalable}. The parameters of the qubit are summarized in \cref{tab:experiment_parameter_field}.
\begin{table}
\caption{Parameters of the superconducting transmon qubit used in the experiments: the qubit frequency $\omega_\mathrm{q}$,  anharmonicity $\alpha$, energy relaxation time $T_1$, and Ramsey dephasing time $T_2^\mathrm{*}$.}
\begin{ruledtabular}
\begin{tabular}{ccccc}
    & $\omega_\mathrm{q}/2\pi$   & $\alpha/2\pi$         & $T_1$                 & $T_2^\mathrm{*}$ \\ \hline
Qubit  & $8.403~{\rm GHz}$ & $-416~{\rm MHz}$    & $20.2~{\rm \mu s}$   & $3.1~{\rm \mu s}$  
\end{tabular}
\end{ruledtabular}
\label{tab:experiment_parameter_field}
\end{table}

In experiments, the quantum gates and measurements are noisy, and thus we do not have perfect implementations of the above gates as mentioned in~\cref{eq:CZ_full_decomp}. Thus, hereafter we use the superscript ``exp" to denote a quantum operation realized in the experiments and hence containing the errors. Let noisy quantum channels that correspond to $Z$, $\mathcal{R}_{Z}\qty(\pm \theta)$ and $\Pi_{\pm Z}$ be $Z^{\rm exp}$, $\mathcal{R}_{Z}^{\rm exp}\qty(\pm \theta)$ and $\Pi^{\rm exp}_{\pm Z}$, respectively. To implement $\mathcal{R}_{Z}^{\rm exp}\qty(\pm \theta)$ and $Z^{\rm exp}$, we use the efficient $Z$-gate technique~\cite{McKay_Efficient_Z_PRA_2017}, in which the rotation around the $Z$-axis can be executed by adjusting the relative phase of the subsequent $X$ and $Y$ qubit drive pulses, which is controlled through the classical hardware and software and can be implemented almost perfectly. The $\mathcal{R}_{X}^{\rm exp}\qty(\pi/2), \mathcal{R}_{Y}^{\rm exp}\qty(\pi/2)$ gates are implemented with the shaped microwave pulses~\cite{half_drag_Lucero_PRA}. For suppressing leakage, we use the DRAG pulsing technique~\cite{Motzoi_drag,Chen_PRL_leakage_suppression}. To reduce the unitary control errors, we use the ORBIT technique~\cite{orbit_PRL}. The average gate fidelity of the single-qubit Clifford gates is $0.9992 \pm 0.0004$ via randomized benchmarking~\cite{RB_Knill,RB_Emerson_IOP_2005,RB_Magesan_PRL_2011}.

To implement $\Pi^{\rm exp}_{\pm Z}$, we perform dispersive readout~\cite{DispersiveReadout_Blais_PRA} followed by classical post-processing. The qubit state is measured through an off-resonantly coupled resonator with the dispersive readout performed via the readout resonator at $10.310~{\rm GHz}$. The readout signal is amplified with an impedance-matched Josephson parametric amplifier~\cite{impa_Mutus_APL,urade2020impa}. The averaged assignment fidelity for the qubit readout is $ 0.9609 \pm0.0037$~(see more details in Appendix.~\ref{app_sec:readout_characterization}). 

\begin{figure*}[t]
\includegraphics[width=\textwidth]{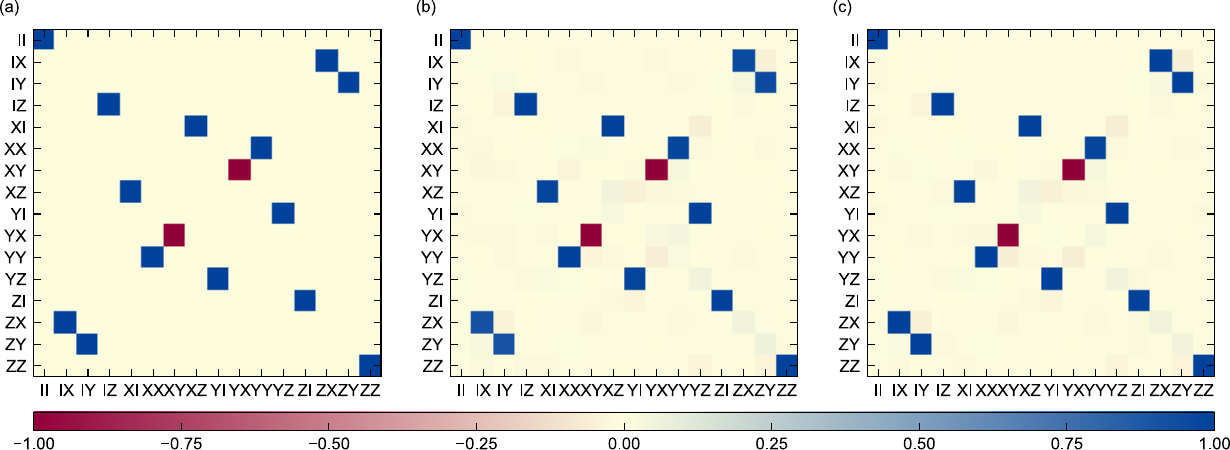}
\caption{\label{fig:virtual_CZ_ptm_results}\textbf{Quantum process tomography results for a CZ gate.} (a) Ideal CZ gate in the Pauli transfer matrix representation. (b) Experimentally-implemented virtual CZ gate without quantum error mitigation. (c) Experimentally-implemented virtual CZ gate with quantum error mitigation.}
\end{figure*}

\subsection{\label{subsec:qpt}Characterization and results}
 To represent quantum states, quantum channels, and measurements, we use the Pauli transfer matrix~(PTM) representation. We define $\sigma_k$~($k=0,\ldots,3$) as the $k$th operator in the Pauli basis, $\mathcal{P} = \{I,X,Y,Z\}$. We define a quantum channel corresponding to a decomposition operation $\mathcal{D} \in \mathcal{D}^{i}_{j}$ as $\Tilde{\mathcal{D}}$. The PTM representation of a quantum channel $\Tilde{\mathcal{D}}$ can be written as $T$, whose elements are given by
\begin{align}
    T(\Tilde{\mathcal{D}})_{k,l} = 
    \frac{1}{d} \mathrm{Tr}\qty[\sigma_{k} \Tilde{\mathcal{D}} (\sigma_{l})] \;,
    \label{eq:ptm_general}
\end{align}
where $d=2^n$ is the dimension of the $n$-qubit system. For the virtual CZ gate characterization, we perform QPT. \CrefSubFigRef{fig:CZ_decomposition}{a} also represents the circuit for virtual two-qubit gate QPT, where we perform QPT in each sub-circuit for the corresponding decomposition channel $\Tilde{\mathcal{D}}_{j=1,\ldots, 5}^{i=\rm A, B}$. As per the QPT procedure, we prepare $d^2=4$ linearly independent states from the basis $\rho_{\rm in}^{i = \rm A,B} = \qty{ \ket{0}, \ket{1}, \ket{+}, \ket{i+}}$, each of which is subjected to the quantum channel $\Tilde{\mathcal{D}}^{i}_{j}$, followed by the quantum state tomography~(QST)~\cite{nielsen_chuang_2010}, which includes measurements of the $\qty{X,Y,Z}$ observables. Here, we have denoted $\ket{\pm}, \ket{i\pm},$ and $\ket{0}$ and $\ket{1}$ as the eigenvectors of $X$, $Y$, and $Z$ with eigenvalues $\pm 1$, respectively. Since we are interested in the QPT of $\Tilde{\mathcal{D}}^{i}_{j}$, we take corresponding $U^{i=\rm A,B}_{\rm pre, post}$ to be the identity gates $I^{\mathrm{A}}$ and $I^{\mathrm{B}}$. These set of experiments constitute a QPT circuit evaluation for the channel $\Tilde{\mathcal{D}}^{i}_{j}$. After obtaining $T(\Tilde{D}_{j}^{i})$ for both sub-circuits A and B, we use~\cref{eq:CZ_full_decomp} to calculate the PTM for the virtual CZ gate $T_{\textrm V-CZ}$. See~\Cref{app_sec:qpt}.

We present the characterization results for the virtual CZ gate and compare the cases with~(PEC) and without~(non-PEC) quantum error mitigation applied for the projection gates in~\cref{fig:virtual_CZ_ptm_results}. For each QPT circuit evaluation of $\Tilde{\mathcal{D}}^{i}_{j}$, we take $N_{\mathrm s}=10,000$ shots to calculate the average expectation value for the corresponding observables $\mathcal{M}_{k=\qty{X,Y,Z}}$. Note that QPT assumes error-free state preparation and measurement~(SPAM). To mitigate initialization errors, we implement initialization by measurement and postselection, in which a measurement pulse is applied at the beginning of each circuit and postselecting the shots which were prepared in the ground state. To mitigate the measurement errors, we apply error mitigation to the measurements, even in experiments without PEC for projection gates, referred to as non-PEC QPT experiments. In contrast, PEC QPT experiments involve the implementation of PEC for both projection gates and the measurements. For the measurement error mitigation in non-PEC QPT experiments, we replace each $\mathcal{M}_{k}$ with the measurement set of Pauli observables $\mathcal{M}_{p} \in \qty{\mathcal{M}_{I},\mathcal{M}_{X},\mathcal{M}_{Y},\mathcal{M}_{Z}}$ as shown in~\crefSubFigRef{fig:CZ_decomposition}{c}. The circuit for PEC QPT experiments is shown in~\crefSubFigRef{fig:CZ_decomposition}{d}, where PEC is applied sequentially first for the projection gates and then for the measurements. See~\Cref{app_sec:mem} for the details. To get the final mitigated expectation value with its uncertainty, we repeat each QPT circuit 100 times and take its average. The average gate fidelity for the virtual CZ gate without doing the quantum error mitigation for the projection gates is $f_{\rm av} = 0.9782 \pm 0.0001$, and after applying the quantum error mitigation to the projection gates it is improved to $f_{\rm av}^{\rm mit} = 0.9938 \pm 0.0002$, where ``mit'' denotes the  fidelity after applying the PEC mitigation. The results indicate a feasible and practical approach to experimentally implement virtual two-qubit gates with enhanced fidelity by employing measurement error mitigation techniques for the projection gates performed through mid-circuit measurements and implementing high-fidelity local single-qubit gates. The high-fidelity virtual two-qubit gates are crucial for simulating quantum circuits using fewer qubits and also to simulate two-qubit gates on a distant pair of qubits which otherwise would take several number of SWAP operations between them.

\section{\label{sec:Conclusion_Discussion}Conclusion and Discussion}
In this work, we experimentally demonstrated a virtual CZ gate and characterized its performance using QPT. Our approach involved formulating probabilistic error cancellation for the projection gates, implemented through mid-circuit measurements. This effectively mitigated projection gate errors. Furthermore, incorporating measurement error mitigation led to a significant enhancement in the average gate fidelity of the virtual CZ gate, even in the presence of higher measurement errors compared to single-qubit gate errors. In one of the applications of the virtual two-qubit gates, where we broke a two-qubit circuit into two disconnected one-qubit circuits, we reduced the number of circuit evaluations for each one-qubit circuit from six to five. 

Decomposing a larger quantum circuit into smaller circuits can boost the capabilities of the limited-sized NISQ devices which can pave the way towards the goal of demonstrating the quantum advantage. However, decomposing a quantum circuit requires large overhead in terms of additional number of circuit evaluations and classical post-processing. These costs vary depending on the decomposition techniques and the requirement of added classical resources. It would be interesting to explore a general approach to decompose a quantum circuit optimally which can also be implemented efficiently in the experiments.

\section*{\label{sec:Acknowledgements}Acknowledgements}
We acknowledge Suguru Endo, Jesper Ilves, Takanori Sugiyama, and Shuhei Tamate for fruitful discussions. This work was partly supported by MEXT Q-LEAP (No.~JPMXS0118068682), JST PRESTO (No.~JPMJPR1916 and No.~JPMJPR2015), JST Moonshot R\&D (No.~JPMJMS2061 and No.~JPMJMS2067), JST ERATO (No.~JPMJER1601), JSPS KAKENHI (No.~JP22H05000 and No.~JP22K17868), MEXT Q-LEAP (No.~JPMXS0118067394), and JST COI- NEXT (No.~JPMJPF2014)

\appendix
\renewcommand\thefigure{\thesection.\arabic{figure}}   
\setcounter{figure}{0} 
\section{Quantum error mitigation}\label{app_sec:mem}
\setcounter{figure}{0} 
\renewcommand\thefigure{A\arabic{figure}}
\begin{figure*}[t!]
\centering
\captionsetup{justification  = RaggedRight, singlelinecheck = false}
\includegraphics[width=\textwidth,keepaspectratio]{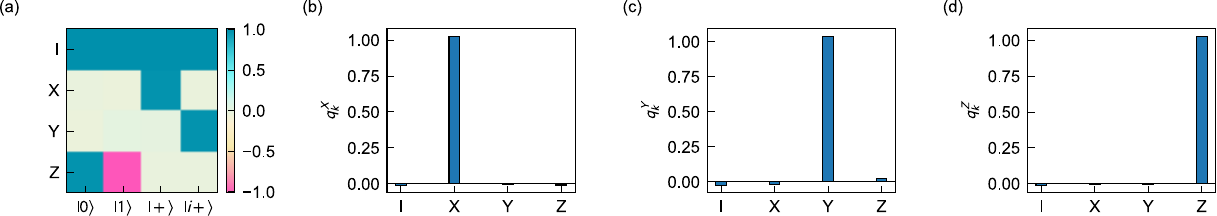}
\caption{\label{fig:g_matrix_qVector}\textbf{Gram matrix and quasi-probability vectors.} (a) Gram matrix for a single-qubit measurement. (b)--(d) Quasi-probability vectors for measurement observables $X$, $Y$, and $Z$, respectively.}
\end{figure*}

We use the probabilistic error cancellation method for QEM~\cite{Endo_PRX_2018,QPMEM_science}. In the PTM representation, a state $\rho$ can be expressed as a column vector 
\begin{align}
    \kket{\rho} = \qty[\rho_0, \rho_1, \cdots , \rho_k, \cdots]^{\rm T} \;, 
\end{align}
whose $k$th element is given by
\begin{align}
    \kket{\rho}_k = \mathrm{Tr}\qty[\rho \sigma_k] \;,
\end{align}
where $\sigma_k$ is defined as the $k$th operator of the Pauli basis $\mathcal{P} = \qty{I,X,Y,Z}$. Similarly, an observable $Q$ can be expressed as a row vector 
\begin{align}
    \bbra{Q} = \qty[Q_0, Q_1, \cdots , Q_k, \cdots] \;,
\end{align}
where 
\begin{align}
    \bbra{Q}_k = \mathrm{Tr}\qty[Q \sigma_k] \;.
\end{align}
A quantum channel $\Tilde{\mathcal{O}}$ corresponding to the operation $\mathcal{O}$ is expressed as 
\begin{align}
    \Tilde{\mathcal{O}}\rho = \mathcal{O} \rho \mathcal{O}^{\dagger} \;,
\end{align}
whose PTM representation is given in~\cref{eq:ptm_general}.

In the QPT experiments, as also mentioned in \cref{subsec:qpt}, we take the initial states as $\rho_{i}= \qty{ \ket{0}, \ket{1}, \ket{+}, \ket{i+}}$ and measure the observables $Q \in \qty{X,Y,Z}$. We want to compute the noise-free (mitigated) expectation values $E^{\rm mit} = \bbraket{Q|\Tilde{O}|\rho}$, however, in experiments we obtain $E^{\rm exp}=\bbraket{Q^{\rm exp}|\Tilde{\mathcal{O}}^{\rm exp}|\rho^{\rm exp}}$. Hence, we apply PEC method to mitigate the errors. The set of initial states can be expressed as the state preparation matrix $A^{\rm exp}$ with its elements $A_{k,i}^{\rm exp} = \bbraket{\sigma_{k} | \rho_{i}^{\rm exp}}$ and the set of observables as the readout matrix $B^{\rm exp}$ with its elements $B_{k,i}^{\rm exp} =  \bbraket{\sigma_{k}^{\rm exp} | \sigma_{i}}$. Since the state preparation errors are much lower in our device, and we also perform initialization by measurement and postselection, we assume the state preparation to be almost perfect~($\rho^{\rm exp} = \rho$) and take its estimate as error-free $A_{k,i}^{\rm exp} \approx \bbraket{\sigma_{k} | \rho_{i}}$, which is a decent guess and is expressed as
\begin{align}
A^{\rm exp} \approx 
    \begin{pmatrix}
        1 & 1 &  1 & 1 \\
        0 & 0 & 1 & 0 \\
        0 & 0 & 0 & 1 \\
        1 & -1 & 0 & 0 \\ 
    \end{pmatrix} \;.
\end{align}
The Gram matrix $\mathbb{G}^{\rm exp} = B^{\rm exp} A^{\rm exp}$, is obtained experimentally by performing the QPT experiment for an identity gate $I$, as also shown in~\crefSubFigRef{fig:g_matrix_qVector}{a}.
With this, we can obtain the readout matrix $B^{\rm exp} = \mathbb{G}^{\rm exp}\qty(A^{\rm exp})^{-1}$.

In the case of unitary decomposition gates, as shown in~\CrefSubFigRef{fig:CZ_decomposition}{a}, the single-qubit gates $\mathcal{O} \in \mathcal{D}^{i=\rm A,B}_{j=\qty{1,2,4,5}}$ are implemented with high-fidelity, so we assume them to be error-free ($\Tilde{\mathcal{O}}^{\rm exp} = \Tilde{\mathcal{O}}$). Due to non-negligible measurement errors, we apply QEM for the measurements, as shown in~\crefSubFigRef{fig:CZ_decomposition}{c}. We can express the observables in terms of experimentally obtained noisy observables~\cite{Temme_qem2017,Endo_PRX_2018} as 
\begin{align}
    \bbra{\sigma_k} = \sum\limits_l q_k^l \bbra{\sigma_l^{\rm exp}} \;,
    \label{app_eq:bbra_sigma_k}
\end{align}
where $k \in \qty{1,2,3}$, $l \in \qty{0,1,2,3}$, $q_k^l$ are the elements of quasi-probability vector $q_k$, and $\bbra{\sigma_l^{\rm exp}}$ is the $l\mathrm{th}$ row of $B^{\mathrm{exp}}$. 
The quasi-probability vectors $q_{k \in \qty{X,Y,Z}}$ are shown in~\crefMultiSubFigRef{fig:g_matrix_qVector}{b)--(d}, respectively, and can be computed as 
\begin{align}
    q_{k} = \bbra{\sigma_{k}} (B^{\rm exp})^{-1} \;.
    \label{app_eq:quasi-prob_vector}
\end{align}
The probability with which $\sigma_l^{\mathrm{exp}}$ is sampled is given by
\begin{align}
    p_{k}^{l}= \frac{|q_{k}^{l}|}{\sum\limits_{m}|q_{k}^{m}|} \;.
    \label{app_eq:quasi-prob}
\end{align}
To obtain the final mitigated expectation value $E^{\rm mit}_k$, each outcome from the experiment is multiplied by a weight factor
\begin{align}
    w_{k}^{l} = \sign (q_{k}^{l})\sum_{m}|q_{k}^{m}| \;.
    \label{app_eq:weight-factor}
\end{align}
The weighted average is the mitigated expectation value for the observable $\sigma_{k}$. Using~\cref{app_eq:bbra_sigma_k,app_eq:quasi-prob_vector,app_eq:quasi-prob,app_eq:weight-factor}, it can be written as
\begin{align}
     E^{\rm mit}_k  &= \sum\limits_{l}q_{k}^{l} \bbra{\sigma_l^{\mathrm{exp}}} \Tilde{\mathcal{D}}^{i}_{j} \kket{\rho} \;, \nonumber \\ 
    & = \sum_{l}w_{k}^{l}p_{k}^{l}\bbra{\sigma_l^{\mathrm{exp}}} \Tilde{\mathcal{D}}^{i}_{j} \kket{\rho} \;,
    \label{app_eq:qpmem_expval_unitary}
\end{align}
where $\Tilde{\mathcal{D}}^{i}_{j} \in \Tilde{\mathcal{D}}^{i=\rm \qty{A,B}}_{j=\qty{1,2,4,5}}$.
In the case of $\expval{\sigma_{0}^{\mathrm{exp}}} = \expval{I^{\mathrm{exp}}}$, we always consider the expectation value as +1. 

\begin{figure*}[t!]
\centering
\captionsetup{justification  = RaggedRight, singlelinecheck = false}
\includegraphics[width=\textwidth,keepaspectratio]{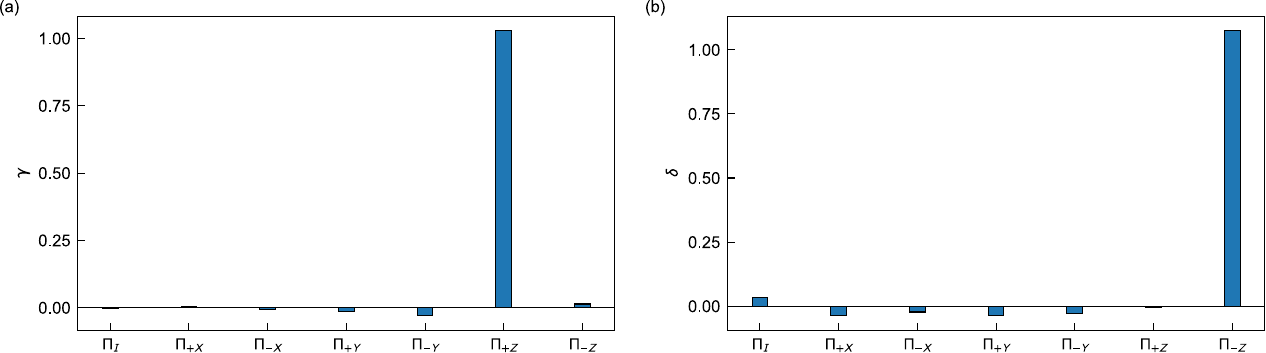}
\caption{\label{fig:proj_q_vec}\textbf{Quasi-probability vectors for projection gates.} (a) Quasi-probability vector $\gamma$ corresponding to projection gate~$\Tilde{\Pi}_{+Z}$. (b) Quasi-probability vector $\delta$ corresponding to projection gate~$\Tilde{\Pi}_{-Z}$.}
\end{figure*}

Now, we discuss applying PEC method on the decomposition circuits containing projection gates $\Pi_{\pm Z}$. We perform the PEC method sequentially first for the projection gates and then for the measurements as shown in \crefSubFigRef{fig:CZ_decomposition}{d}. Since we implement projection gates through mid-circuit measurements, we make an approximation here, i.e., we express a projection gate operation~($\Tilde{\Pi}_{\pm Z}$) as a linear combination of the basis operations $\mathcal{B}^{\rm exp} \in \qty{\Tilde{I}, \Tilde{\Pi}_{+X},\Tilde{\Pi}_{-X},\Tilde{\Pi}_{+Y},\Tilde{\Pi}_{-Y},\Tilde{\Pi}_{+Z},\Tilde{\Pi}_{-Z}}$. The PTMs $\mathcal{B}^{\rm exp}$ are estimated using~\cref{eq:ptm_general} with measurement error mitigation. We can express the PTMs corresponding to $\Pi_{Z}$ as
\begin{align}
    \Tilde{\Pi}_{+Z} \simeq \sum\limits_{u} \gamma_{u} \mathcal{B}_u^{\rm exp} \;,
    \label{app_eq:pi_+z}
\end{align}
\begin{align}
    \Tilde{\Pi}_{-Z} \simeq \sum\limits_{u}  \delta_{u} \mathcal{B}_u^{\rm exp} \;,
    \label{app_eq:pi_-z}
\end{align}
where $u \in \qty{1,\ldots,7}$ denotes the $u$th element of $\mathcal{B}^{\rm exp}$, and $\gamma$ and $\delta$ are quasi-probability vectors, as shown in~\cref{fig:proj_q_vec}. To calculate $\gamma$ and $\delta$, we minimize the Euclidean 2-norms $||\sum\limits_{u} \gamma_{u} \mathcal{B}_u^{\rm exp} -  \Tilde{\Pi}_{+Z}||$ and $||\sum\limits_{u} \delta_{u} \mathcal{B}_u^{\rm exp} -  \Tilde{\Pi}_{-Z}||$, respectively. 

The probabilities of selecting $\mathcal{B}_u^{\rm exp}$ corresponding to $\Tilde{\Pi}_{+Z}$  and $\Tilde{\Pi}_{-Z}$ from~\cref{app_eq:pi_+z,app_eq:pi_-z} can therefore be expressed as
\begin{subequations}
\label{app_eq:prob_projection}
    \begin{align}
p_{u}^\gamma = \frac{|\gamma_{u}|}{\sum\limits_{r}|\gamma_u^{r}|} \;, 
\label{app_eq:prob_proj_+z}\\
p_{u}^\delta = \frac{|\delta_{u}|}{\sum\limits_{s}|\delta_u^{s}|} \;,
\label{app_eq:prob_proj_-z}
\end{align}
\end{subequations}
respectively. The respective weight factors associated with \cref{app_eq:prob_proj_+z,app_eq:prob_proj_-z} can be written as 
\begin{subequations}
\label{app_eq:weight_projection}
    \begin{align}
     w_{u}^\gamma = \sign (\gamma_{u})\sum_{r}|\gamma_{u}^{r}| \;, \\
     w_{u}^\delta = \sign (\delta_{u})\sum_{s}|\delta_{u}^{s}| \;.
\end{align}
\end{subequations}

We can express the mitigated expectation value of observables $\sigma_k^\prime$ as
\begin{align}
    E_{k^\prime}^{\rm mit} = \bbra{\sigma_{k^\prime}} \Tilde{\mathcal{D}}^{i}_{3} \kket{\rho} \;,
    \label{app_eq:mit_exp-val_relation_proj}
\end{align}
where $k^\prime = \qty{1,2,3}$ and $\Tilde{\mathcal{D}}^{i=\qty{\rm A,B}}_{3} = \Tilde{\Pi}_{\pm Z}$. Using~\cref{app_eq:bbra_sigma_k,app_eq:quasi-prob_vector,app_eq:quasi-prob,app_eq:weight-factor} and \cref{app_eq:pi_+z,app_eq:pi_-z,app_eq:prob_projection,app_eq:weight_projection}, we can express \cref{app_eq:mit_exp-val_relation_proj} for $ \Tilde{\Pi}_{+ Z}$ and $ \Tilde{\Pi}_{-Z}$ as
\begin{subequations}
    \begin{align}
         E^{\rm mit}_{+Z, k^\prime} = 
         \sum_{l} \sum_{u}
         w_{k^\prime}^{l}p_{k}^{l}w_u^\gamma p_u^\gamma 
         \bbra{\sigma_l^{\mathrm{exp}}}\mathcal{B}_u^{\rm exp} \kket{\rho} \;, \\
         E^{\rm mit}_{-Z, k^\prime} = 
         \sum_{l} \sum_{u}
         w_{k^\prime}^{l}p_{k}^{l}w_u^\delta p_u^\delta 
         \bbra{\sigma_l^{\mathrm{exp}}} \mathcal{B}_u^{\rm exp} \kket{\rho} \;,
    \end{align}
\end{subequations}
respectively.

\section{Qubit readout characterization}\label{app_sec:readout_characterization}
\setcounter{figure}{0} 
To obtain the average assignment fidelity of the single-shot readout, we apply a $\mathcal{R}_X(\pi/2)$ gate on the initial ground state $\ket{0}$ and then apply two sequential measurement pulses. The average assignment fidelity is defined as

\begin{align}
    \mathcal{F}_{a} = \frac{1}{2} \Bigl( p(g|g) + p(e|e) \Bigr),
\end{align}
where $p(x|y)$ is the probability of assigning the measurement outcome $x$ when the qubit is prepared in the state $y$.
We obtain the assignment fidelity of $0.9609 \pm 0.0037$ and the ground-state initialization fidelity $\mathcal{F}_{\rm init} = p(g|g)$ of $0.9895 \pm 0.0028$ . 

\section{Quantum process tomography}\label{app_sec:qpt}
\setcounter{figure}{0} 
We use QPT to evaluate the performance of the virtual two-qubit gate. In general, for a $n$-qubit system with the dimension $d=2^{n}$, we can express a quantum channel $\Tilde{\mathcal{E}}$ acting on an arbitrary quantum state $\rho$ as
\begin{align}
    \Tilde{\mathcal{E}} \qty(\rho)  = \sum_{i,j=0}^{d^{2}-1} \chi_{ij}B_{i}\rho B_{j}^{\dagger},
    \label{app_eq:qpt}
\end{align}
where $\qty{B_{i}}$ are the elements of the $d \times d$ matrix basis and $\chi$ is the process matrix~\cite{nielsen_chuang_2010}. We define the PTM representation of a quantum channel $\Tilde{\mathcal{E}}$ as $T$, whose elements are expressed as
\begin{align}
    T(\Tilde{\mathcal{E}})_{i,j} = \Tr\qty[\sigma_{i} \Tilde{\mathcal{E}}( \sigma_{j})].
    \label{app_eq:ptm}
\end{align}
In the PTM representation, we can express a state $\rho$ as a column vector $\ket{\rho}\rangle$ with elements $\ket{\rho}\rangle_{k} = \Tr\qty[\sigma_{k}\rho]$ and an operator $\mathcal{O}$ as a row vector $\langle \bra{\mathcal{O}}$ with elements $\langle \bra{\mathcal{O}}_{k} = \Tr\qty[\sigma_{k}\mathcal{O}]$. In the QPT, we prepare for each qubit an initial state from the set $\qty{ \ket{0}, \ket{1}, \ket{+}, \ket{i+}}$. For each initial state, we measure the qubit along the Pauli basis, $X,Y,$ and $Z$. 

Using the PTM representation, we can define the quantum channel for the two-qubit system as the tensor product of individual PTMs. Using~\cref{eq:CZ_full_decomp,eq:ptm_general}, we can express the PTM for the virtual CZ gate $T_{\textrm V-CZ}$ as
\begin{widetext}
\begin{align}
    T_{\textrm V-CZ} &= \frac{1}{2}\sum_{i=1,2}\qty(T(\Tilde{\mathcal{D}}_{i}^{\rm A}) \otimes T(\Tilde{\mathcal{D}}_{i}^{\rm B})) 
     -\frac{1}{2} \sum_{\alpha_1, \alpha_2,\beta}\alpha_1, \alpha_2
    \Bigg[
    \qty(T(\Tilde{\mathcal{D}}_{\alpha_1 3}^{\rm A}) \otimes T(\Tilde{\mathcal{D}}_{\beta}^{\rm B})) 
    + \qty(T(\Tilde{\mathcal{D}}_{\beta}^{\rm A}) \otimes T(\Tilde{\mathcal{D}}_{\alpha_2 3}^{\rm B}))
    \bigg],
\end{align}
\end{widetext}
where $\alpha_{1}, \alpha_{2} \in \{\pm 1 \}$ and $\beta \in \{4,5\}$. 
The average gate fidelity in terms of PTMs is defined as 
\begin{align}
    f_{\mathrm{av}}(\Tilde{\mathcal{U}},\Tilde{\mathcal{E}}) = \frac{1}{d+1}\qty[\frac{1}{d}\sum\limits_{i,j}T(\Tilde{\mathcal{U}})_{i,j}T(\Tilde{\mathcal{E}})_{i,j}+1],
\end{align}
where $d=2^n$ is the dimension of the $n$-qubit system~\cite{heya_vqgo}. $T(\Tilde{\mathcal{U}})$ is the PTM of the ideal target gate~($T_{CZ}$ for CZ gate), and $T(\Tilde{\mathcal{E}})$ is the PTM of experimentally implemented gate~($T_{\textrm V-CZ}$ for virtual CZ gate in this case).

\end{document}